\documentclass[12pt]{article}

\usepackage{graphicx}
\usepackage{times}

\title{Subtree power analysis finds optimal species for comparative genomics}

\author
{Jon D. McAuliffe,$^{1}$ Michael I. Jordan,$^{1,2}$ Lior Pachter$^{3,\ast}$ \\
\\
\normalsize{$^{1}$Department of Statistics,}
\normalsize{$^{2}$Computer Science Division,}
\normalsize{$^{3}$Department of Mathematics,} \\
\normalsize{University of California,} \\
\normalsize{Berkeley, CA 94720, USA} \\
\\
\normalsize{$^\ast$To whom correspondence should be addressed;
            E-mail:  lpachter@math.berkeley.edu.}
}

\date{}

\newcommand{\myfig}[1]{Figure~\ref{fig:#1}}
\newcommand{\mytbl}[1]{Table~\ref{tbl:#1}}

\newcommand{\eqref}[1]{(\ref{eqn:#1})}

\begin{document}

\maketitle

\begin{abstract}
  Sequence comparison across multiple organisms aids in the detection of
  regions under selection. However, resource limitations require a
  prioritization of genomes to be sequenced. This prioritization should be
  grounded in two considerations: the lineal scope encompassing the
  biological phenomena of interest, and the optimal species within that
  scope for detecting functional elements. We introduce a statistical
  framework for optimal species subset selection, based on maximizing power
  to detect conserved sites. In a study of vertebrate species, we show that
  the optimal species subset is not in general the most evolutionarily
  diverged subset. Our results suggest that marsupials are prime sequencing
  candidates.
\end{abstract}

\section*{Introduction}

Comparative genomic methods can reveal conserved regions in multiple
organisms, including functional elements undetected by single-sequence
analyses~\cite{WatEtAl2002,Rat2004}. Individual studies have demonstrated
the effectiveness of genomic comparison for specific regions and
elements~\cite{Fli2001,BofEtAl2003,DerEtAl2003,ThoEtAl2003,ChaEtAl2004}.
Such successes indicate that comparative considerations should play a major
role in decisions about what unsequenced species to sequence next. For
comparative purposes, sequencing choices must first of all be guided by
specification of the widest range of species sharing the functions or
characters in question, which we call the lineal scope~\cite{OBrEizMur2001}
\footnote{This differs from the ``phylogenetic scope'' of Cooper \textit{et
    al.} \cite{CooEtAl2003} in that lineal scope is determined solely by a
  biological trait of interest, whereas phylogenetic scope can be
  determined according to any considerations.}. Boffelli \textit{et
  al.}~\cite{BofNobRub2004} discuss the utility of comparisons in lineal
scopes ranging from the primate clade to the vertebrate tree.

Most lineal scopes selected in practice will include far more extant
species than can be sequenced with today's resources. Thus, sequencing
prioritization is an unavoidable issue, both for smaller-scale efforts
targeting particular regions and for whole-genome projects, whose focus
should reflect in part the aggregate needs of comparative analyses. Few
studies on comparative methods provide a quantitative framework for
decision-making about what to sequence. An exception is the work of Sidow
and others~\cite{CooEtAl2003,Sid2002}: given a set of sequenced organisms
and an inferred phylogeny, Cooper \textit{et al.}~\cite{CooEtAl2003} argue
that decisions should be based on maximizing additive evolutionary
divergence in a phylogenetic tree.

While additive divergence captures part of the problem underlying organism
choice, it fails to reflect the inherent tradeoff that characterizes the
problem. On the one hand, the success of procedures for assessing
conservation does depend on sufficient evolutionary distance among the
sequences~\cite{DerEtAl2003,BofEtAl2003,MarEtAl2003}. On the other hand, a
given set of species may have diverged too far from one another to be
useful, even when orthology is preserved: in the limit of large
evolutionary distance, conservation and nonconservation are just as
indistinguishable as at distance zero~\cite{ZhaEtAl2003}. Furthermore,
phylogenetic topology has counterintuitive effects on usefulness.

Here, we present a decision-theoretic framework which subsumes these
issues, providing a procedure for making systematic, quantitative choices
of species to sequence. Statistical power is our optimality criterion for
species selection. Thus, we measure the effectiveness of a species subset
directly in terms of error rates for detecting and overlooking conservation
at a single orthologous site. Measuring power disentangles effects due to
the number of species used from effects due to relative evolutionary
distances in the phylogeny. We illustrate these ideas theoretically, in a
star phylogeny analysis, and practically, with an empirically-derived
phylogeny on 21 representative vertebrate species. The results indicate
that adding the dunnart or a closely-related marsupial to finished and
underway vertebrate sequences would most increase the power to detect
conservation at single-nucleotide resolution.

\section*{Setup}

We frame conservation detection in the following decision-theoretic
setting. The data $\mathbf{x}$ are the nucleotides at an orthologous site
across a set of species, i.e., an ungapped alignment column. We view these
bases as corresponding to the leaves of a phylogeny with unobserved
ancestral bases. We assume that the phylogenetic topology, the Markov
substitution process along the branches, and the branch lengths are all
known. The phylogeny induces the observed-data probability distribution
$p(\mathbf{x}; r)$ as the marginal distribution on its leaves, which can be
evaluated efficiently for any $\mathbf{x}$ and $r$~\cite{Fel1981}. The
parameter $r > 0$ is an unknown global mutation rate shared among all
branches. We choose two threshold values $r_N > r_C$ for $r$: an actual
mutation rate of at least $r_N$ corresponds by definition to a nonconserved
site, whereas a rate no more than $r_C$ means the site is strongly
conserved. When $r_N > r > r_C$, the conservation is too weak to interest
us.

The decision-theoretic goals are now twofold. First, fixing a set of
species, we wish to select a decision rule $\delta(\mathbf{x})$ which
declares the site either nonconserved ($\delta(\mathbf{x}) = 0$) or
conserved ($\delta(\mathbf{x}) = 1$) using only data from those species.
Every nontrivial $\delta(\mathbf{x})$ will have positive probability of
making two mistakes: when $r \geq r_N$, $P_r(\delta(\mathbf{X}) = 1)$ is
the probability it erroneously detects conservation, and when $r \leq r_C$,
$P_r(\delta(\mathbf{X}) = 0)$ is the probability it overlooks conservation.
Minimizing these probabilities guides our choice of $\delta(\mathbf{x})$.
We formulate a Neyman-Pearson hypothesis test~\cite{Leh1986} of the null
hypothesis \mbox{$H_0$: $r \geq r_N$} versus the alternative
hypothesis \mbox{$H_A$: $r \leq r_C$}, stipulating a maximum allowed
probability $\alpha$ of falsely rejecting $H_0$ (falsely declaring
conservation). Controlling this error probability is a central
concern~\cite{CooEtAl2003}. Subject to this constraint, we find a test
statistic $\delta(\mathbf{x})$ with large power to detect conservation,
that is, small probability of overlooking conservation. The second goal is
to maximize this power over subtrees in the larger phylogeny determined by
the chosen lineal scope, such as all subtrees on $k$ extant species within
the anthropoid clade, where $k$ is determined by sequencing resource
limitations.

\section*{Symmetric star topology}

\begin{figure}
  \centering
  \includegraphics[scale=0.97]{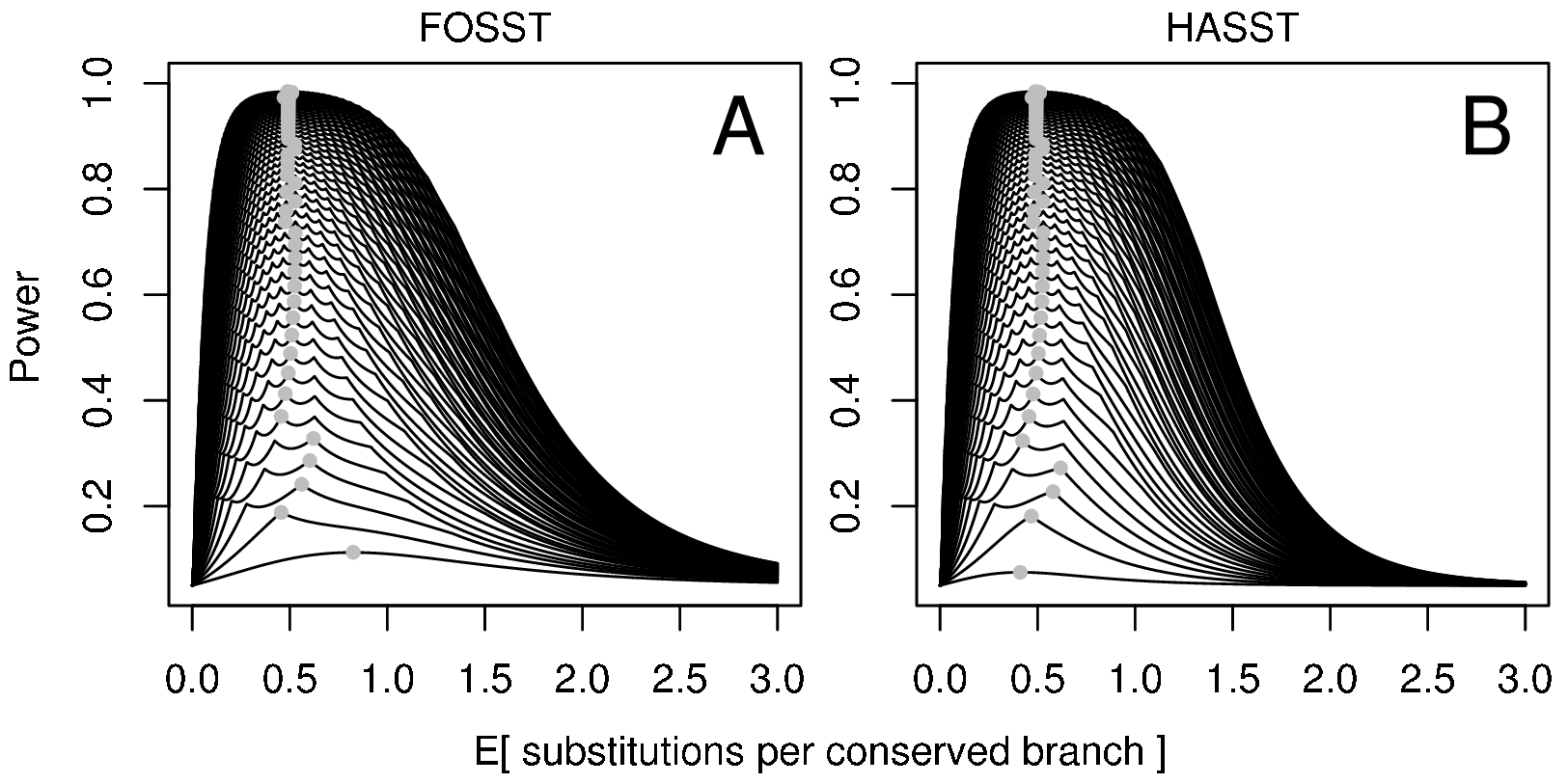}
  \caption{Power to detect conservation as a function of common branch
    length for the fully-observed (A) and hidden-ancestor (B)
    \textsc{sst}s, using $r_C = 1$, $r_N = 2$, and $\alpha = 0.05$.  Each
    power curve corresponds to an even number $k$ of observed descendant
    species, from two (bottommost curve) to 100 (topmost). The maximum
    power attained for each $k$ is indicated by a grey dot. The power
    analysis uses the Jukes-Cantor substitution process; power curves are
    computed analytically (see Appendix).  Power curves computed with
    other values of $r_N$ and $\alpha$ remain qualitatively the same (not
    shown).}
  \label{fig:sstpower}
\end{figure}

We initially pursue these goals in a phylogenetic setting called the
symmetric star topology (\textsc{sst}), where $k$ extant species are
connected to a single ancestor by branches of common length $t > 0$.
Choosing $k$ and $t$ in the \textsc{sst} is akin to choosing $k$ extant
species within a larger phylogeny, such that each pair of chosen species is
at a distance of approximately $2t$. Hypothesis testing in the
fully-observed \textsc{sst} (\textsc{fosst}), with known ancestral base,
closely approximates testing in the hidden-ancestor \textsc{sst}
(\textsc{hasst}), the case of interest, for small to moderate $t$
(\myfig{sstpower}). This follows because there is little uncertainty about
the ancestral base at short evolutionary distances: with high probability,
it equals the most-occurring base among the descendants. The analogy
matters because we know the uniformly most-powerful testing procedure under
the \textsc{fosst}: it rejects $H_0$ (declares conservation) for large
values of the likelihood ratio statistic \mbox{$p(\mathbf{x}; r_C) /
  p(\mathbf{x}; r_N)$} (see Appendix).

\begin{figure}
  \centering
  \includegraphics[scale=0.97]{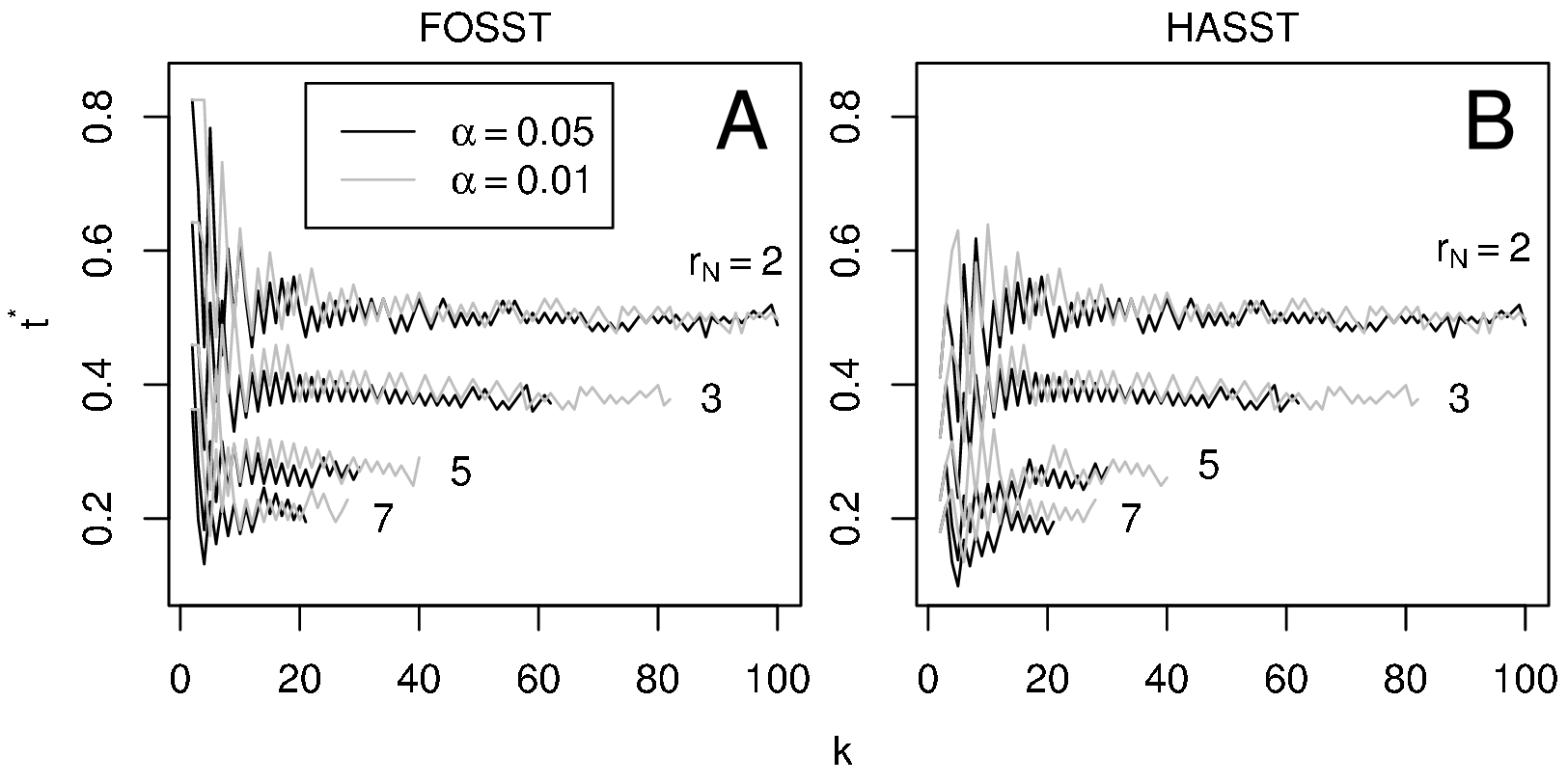}
  \caption{The optimal common branch length $t^*(k)$ in the fully-observed
    (A) and hidden-ancestor (B) \textsc{sst}s, as a function of the
    number of descendant species $k$. Each black curve uses the indicated
    nonconserved rate \mbox{$r_N = 2,3,5,7$} with $\alpha = 0.05$; grey
    curves are analogous with $\alpha = 0.01$. As $k$ increases, $t^*(k)$
    stabilizes at a value depending on $r_N$ but not $\alpha$. For the
    larger $r_N$'s, the curves are terminated when power reaches 99.9\%.}
  \label{fig:ssttstark}
\end{figure}

\myfig{sstpower}A shows the power of the \textsc{fosst} likelihood-ratio
test against the particular alternative distribution $r = r_C$, as $t$ and
$k$ vary. Power against other alternatives $r < r_C$ is larger (see
Appendix). For each $t$, power increases monotonically in $k$. However, for
each $k$, there is a unique power-maximizing branch length $t^*(k)$. In the
Appendix we explain this in terms of stationary Markov substitution
processes. Fundamentally, it happens because both nonconserved and
conserved sites accrue mutations, and the difference in their mutation
rates becomes irrelevant as $t \to \infty$. A consequence of this is the
suboptimality of maximizing additive divergence: for any $k$, the optimal
tree has finite divergence $k \cdot t^*(k)$, rather than arbitrarily large
divergence. Comparing Figures~\ref{fig:sstpower}A and \ref{fig:sstpower}B
shows the \textsc{fosst} accurately approximates the \textsc{hasst} in a
large interval around $t^*(k$) for $k > 2$, so the conclusion also applies
to the \textsc{hasst}. As $k$ increases, $t^*(k)$ stabilizes at a nonzero
value (\myfig{ssttstark}). Thus, the optimal divergence $k \cdot t^*(k)$
grows without bound as a function of $k$.

\section*{Empirical power analysis}

We now explore subtree power maximization empirically, using the
previously-reported CFTR sequence data~\cite{ThoEtAl2003} on 21
representative vertebrates~(\mytbl{species}). We estimated a phylogeny
(\myfig{tree}) based on a multiple sequence alignment, as described in the
Appendix. This procedure yields phylogenies applicable to data outside the
estimation region~\cite{CooEtAl2003,YapPac2004}. We formulated the
likelihood-ratio statistic and calibrated the conserved rate threshold
$r_C$ to correspond to typical genic conservation in the sequenced region.
Having fixed the form of the testing procedure, the goal is to maximize its
power to detect conservation over subsets of size $k$ chosen from among the
21 species, for various values of $k$. This entails searching for the
maximal-power family subtree, or $k$-most-powerful Steiner subtree
(k-\textsc{mpss}), among the ${21 \choose k}$ subtrees with $k$ leaves (see
Appendix). A Steiner subtree on $k$ leaves is the unique smallest subtree
rooted at their last common ancestor.

\mytbl{stp} shows the $k$-\textsc{mpss} (starred) in comparison to the subtree
on $k$ leaves with largest additive divergence (the $k$-most-divergent
Steiner subtree, or $k$-\textsc{mdss}, daggered). The latter has been the
focus of previous work~\cite{BofEtAl2003,McaPacJor2004,CooEtAl2003}. These
two subtree selection criteria do not coincide. For instance, at $r_N = 2$,
the 5-\textsc{mpss} includes the dunnart, whereas the 5-\textsc{mdss}
instead uses the platypus. The $t$-statistic on the difference in power is
$2.06$, so variability in the power estimate is not a likely explanation.
A more extreme example is $r_N = 10$: the 4-\textsc{mpss} and
4-\textsc{mdss} have only one species in common, and the absolute loss in
power that results from using the 4-\textsc{mdss} is nearly 8.5\%
($t$-statistic 105.7). Here, more than 4,400 subtrees have higher power
than the 4-\textsc{mdss}. The disagreement at higher values of $k$
underscores the effect of phylogenetic topology on the detection of
conservation.

We carried out a similar comparison, under the constraint that the 9
completely or partially sequenced vertebrates in the data set appear in the
subtree (\mytbl{clampedstp}). This reveals the species whose addition to
the current sequencing mix would most improve the power to detect
single-site conservation. As in \mytbl{stp}, the most-powerful and
most-divergent subtrees generally differ. The pattern of disagreement is
not systematic: when $r_N = 5$, for example, they disagree at 10 and 11
species, agree at 12 and 13, and disagree at 14. \mytbl{stp} exhibits
similar properties. We conclude that the $k$-\textsc{mdss} is not a
reliable surrogate for the $k$-\textsc{mpss}. \mytbl{clampedstp} reveals
that the single most beneficial species to sequence next is the dunnart
(improving power by a relative 12.5\%), whereas the species which adds the
most evolutionary divergence is the platypus.

\section*{Discussion}

Even when the \textsc{mpss} and \textsc{mdss} coincide, the
decision-theoretic point of view puts the focus on the important issue: the
two kinds of discrimination errors and their probabilities. The power
calculation directly measures the marginal benefit of additional sequenced
species as an increase in probability of conserved site detection. This
enables us to choose a $k$ which optimizes the tradeoff between the
expected benefit of detecting conservation and the cost of additional
sequencing. By contrast, the additive divergence of a species set gives no
direct indication of how a procedure using those species will fare. Since
the phylogeny and substitution process are parameters of our procedure,
their choice can be tailored to particular investigations. Our emphasis on
single-site detection of conservation will lead to conservative power
estimates in situations where conservation is tested for simultaneously
across multiple correlated sites.

\section*{Acknowledgments}

We thank Peter Bickel and Adam Siepel for helpful comments. L.P. was
supported by a grant from the NIH (R01-HG2362-3), a Sloan Foundation
Research Fellowship, and an NSF Career award (CCF-0347992).

\section*{Appendix}

\subsection*{Symmetric star topology}

\vspace{.3cm}

\subsubsection*{Fully observed}

Let $\mathcal{P} = \{ p(x_0, \mathbf{x}; r)$ : $r > 0 \}$ be the family of
\textsc{fosst} probability mass functions indexed by the mutation rate
parameter $r$, for some fixed choice of descendant species count $k$ and
common branch length $t$. Here $x_0$ is the observed ancestral base and
$\mathbf{x} = (x_1, \ldots, x_k)$ are the observed descendant bases. Write
\begin{equation}
  n(x_0, \mathbf{x}) = \sum_{i=1}^k \delta(x_0, x_i) \ , \label{eqn:n}
\end{equation}
where $\delta(\cdot,\cdot)$ is the Kronecker delta function. Under the
Jukes-Cantor substitution process, with its equilibrium distribution (the
uniform distribution) on the ancestral base, each member of $\mathcal{P}$
has the form
\begin{eqnarray}
p(x_0, \mathbf{x}; r) & = & \frac{1}{4} \prod_{i=1}^k \left(\frac{1 +
    3e^{-4rt}}{4}\right)^{\delta(x_0,x_i)}
\left(\frac{3(1 - e^{-4rt})}{4}\right)^{1 - \delta(x_0, x_i)} \\
& = &  \frac{1}{4} \left(\frac{1 + 3e^{-4rt}}{4}\right)^{\sum_{i=1}^k
  \delta(x_0,x_i)}
\left(\frac{3(1 - e^{-4rt})}{4}\right)^{k - \sum_{i=1}^k \delta(x_0,x_i)}
\ . \label{eqn:p}
\end{eqnarray}
Fixing $r_C = 1$ entails no loss in generality, due to the
nonidentifiability of the parameter pair $(r,t)$ in the Jukes-Cantor
substitution process. Choose $r_N > 1$. Substituting~\eqref{n}
into~\eqref{p} and simplifying the ratio $p(x_0, \mathbf{x}; 1) / p(x_0,
\mathbf{x}; r_N)$ shows that the likelihood-ratio statistic for testing
\mbox{$H_0$ : $r \geq r_N$} versus \mbox{$H_A$ : $r \leq 1$} in the
\textsc{fosst} model has the form
\begin{equation}
  \frac{(1 + 3e^{-4t})^{n(x_0,\mathbf{x})}(1 - e^{-4t})^{k -
    n(x_0,\mathbf{x})}}
  {(1 + 3e^{-4 r_N t})^{n(x_0,\mathbf{x})}
    (1 - e^{-4 r_N t})^{k - n(x_0,\mathbf{x})}} \ .
  \label{eqn:fosstlrs}
\end{equation}
The family $\mathcal{P}$ has a monotone (decreasing) likelihood ratio in
the statistic $n(x_0, \mathbf{x})$, because for each pair of rate
parameters $r_1 > r_2 > 0$, the likelihood ratio
\begin{equation}
  \frac{(1 + 3e^{-4 r_1 t})^n (1 - e^{-4 r_1 t})^{k - n}}
  {(1 + 3e^{-4 r_2 t})^n (1 - e^{-4 r_2 t})^{k - n}} =
  \left( \frac{1 + 3e^{-4 r_1 t}}{1 + 3e^{-4 r_2 t}} \right)^n
  \left( \frac{1 - e^{-4 r_1 t}}{1 - e^{-4 r_2 t}} \right)^{k - n}
\end{equation}
is a decreasing function of $n = n(x_0, \mathbf{x}) \in \{ 0, 1, \ldots, k
\}$. This follows upon observing that, when $r_1 > r_2$,
\[
\frac{1 + 3e^{-4 r_1 t}}{1 + 3e^{-4 r_2 t}} < 1 \mathrm{\ \ \ \ and\ \ \ \ \,}
\frac{1 - e^{-4 r_1 t}}{1 - e^{-4 r_2 t}} > 1 \ .
\]
Standard monotone likelihood-ratio theory~\cite{Leh1986} therefore implies
that the likelihood-ratio test $\mathcal{T}_\alpha$, which rejects
when~\eqref{fosstlrs} exceeds a critical value $c_\alpha$, is uniformly
most powerful for testing \mbox{$H_0$ : $r \geq r_N$} versus \mbox{$H_A$ : $r
  \leq 1$} at size $\alpha$. The size is attained at the particular null
distribution $r = r_N$.

The theory also implies that, among the alternative distributions in $H_A$,
$\mathcal{T}_\alpha$ attains its lowest power against $r = 1$, yielding a
lower bound on the power against any member of $H_A$. The power of
$\mathcal{T}_\alpha$ against the particular alternative $r = 1$ can be
written explicitly as a function of $k$ and $t$:
\begin{equation}
  \rho(k,t) = G_A(n_\alpha + 1; k) + \left(\frac{\alpha -
      G_0(n_\alpha + 1; k)}{f_0(n_\alpha; k)}\right)
  f_A(n_\alpha; k) \ . \label{eqn:fosstpower}
\end{equation}
Here, $f_0(\cdot;k)$ and $f_A(\cdot;k)$ are the probability mass functions
of a $\mathrm{Bin}(k, d(r,t))$ random variable with $r = r_N$ and $r = 1$,
respectively; $G_0(\cdot;k)$ and $G_A(\cdot;k)$ are the corresponding
(cadlag) cumulative binomial right-tail probabilities; $d(r,t) = (1 +
3\exp(-4rt))/4$; and $n_\alpha$ is a known critical value. To
derive~\eqref{fosstpower}, first note that $\mathcal{T}_\alpha$ is
equivalent to the test which rejects $H_0$ when the statistic $n(x_0,
\mathbf{x})$ exceeds a corresponding critical value $n_\alpha$, again by
virtue of the monotone likelihood-ratio property. Both tests thus have the
same power $\rho(k,t)$. Let $P_0$ and $P_A$ denote the distribution of
$n(X_0,\mathbf{X})$ under $r = r_N$ (the size-determining distribution) and
$r = 1$, respectively. Because $n(x_0, \mathbf{x})$ can take on only
finitely many values, we use randomized rejection to achieve level exactly
$\alpha$. The critical value is $n_\alpha = \min\{n$ : $P_0(n(X_0,
\mathbf{X}) > n) \leq \alpha\}$.  When $n(x_0, \mathbf{x}) > n_\alpha$, we
reject. When $n(x_0, \mathbf{x}) = n_\alpha$, we reject with probability
$\gamma(\alpha)$ satisfying
\begin{equation}
  P_0(n(X_0,\mathbf{X}) > n_\alpha) +
  \gamma(\alpha) P_0(n(X_0, \mathbf{X}) = n_\alpha) = \alpha \ .
\end{equation}
This implies that setting
\begin{equation}
  \gamma(\alpha) = \frac{\alpha - P_0(n(X_0, \mathbf{X}) >
    n_\alpha)}{P_0(n(X_0, \mathbf{X}) = n_\alpha)} \label{eqn:gamma}
\end{equation}
guarantees a test with size $\alpha$. It now follows that
\begin{equation}
\rho(k,t) = P_A(n(X_0,\mathbf{X}) > n_\alpha) +
\gamma(\alpha) P_A(n(X_0, \mathbf{X}) = n_\alpha) \ . \label{eqn:rho}
\end{equation}
Under the star topology and Jukes-Cantor substitution process, each
descendant nucleotide $X_i$ has probability $d(r,t) = (1 + 3\exp(-4rt))/4$
of differing from $X_0$, independent of all other descendants. Thus $n(X_0,
\mathbf{X})$ is a $\mathrm{Bin}(k, d(r,t))$ random variable.
Equation~\eqref{fosstpower} follows upon substituting $G_0(n_\alpha + 1;
k)$ for $P_0(n(X_0,\mathbf{X}) > n_\alpha)$, $f_0(n_\alpha; k)$ for
$P_0(n(X_0, \mathbf{X}) = n_\alpha)$, and similarly for $P_A$.

Equation~\eqref{fosstpower} involves only known constants and binomial
probabilities. The latter can be evaluated quickly to desired
accuracy~\cite{AbrSte1974}. This allows us to compute $\rho(k,t)$ for many
choices of $k$ and $t$, leading to the power curves in \myfig{sstpower}A.
The kinks in each power curve correspond to settings of $t$ at which the
critical value of the likelihood-ratio test changes. The locations of the
kinks are easily determined, and the power curves are highly smooth between
kinks. Thus, we can find $t^*(k)$ and $\rho^*(k)$ rapidly using a numerical
optimization routine (\myfig{sstpower}A, \myfig{ssttstark}A).

\subsubsection*{Hidden ancestor}

Under the \textsc{hasst} model and Jukes-Cantor process, the
likelihood-ratio statistic has the form
\begin{equation}
  \frac{\sum_{x_0} (1 + 3e^{-4t})^{n(x_0,\mathbf{x})}
    (1 - e^{-4t})^{k - n(x_0,\mathbf{x})}}
  {\sum_{x_0} (1 + 3e^{-4 r_N t})^{n(x_0,\mathbf{x})}
    (1 - e^{-4 r_N t})^{k - n(x_0,\mathbf{x})}} \ .
\label{eqn:hasstlrs}
\end{equation}
This is more difficult to deal with than \eqref{fosstlrs}. It is clear that
\eqref{hasstlrs} depends only the occurrence counts of the four different
bases, not on the leaf configuration which gives rise to the counts.
Indeed, \eqref{hasstlrs} is invariant when the bases associated with the
counts are permuted. This means that there are only as many distinct values
of \eqref{hasstlrs} as there are integer partitions of $k$ into four parts,
with partition values of zero allowed. The number of leaf configurations
corresponding to each integer partition is an easy combinatorial quantity.
We can generate all the integer partitions and evaluate the \textsc{hasst}
probability mass function at each one quickly, even for $k$ as large as
100.

Together, these facts allow us to compute the exact null distribution $r =
r_N$ and alternative distribution $r = 1$ of \eqref{hasstlrs}, for each
required setting of $(\alpha, r_N, k, t)$. This yields the power of the
\textsc{hasst} likelihood-ratio test, using formulas~\eqref{gamma}
and~\eqref{rho} with the \textsc{hasst} distribution functions substituted
for $P_0$ and $P_A$. We then maximize each curve $\rho(k,\cdot)$ by brute
force to determine $t^*(k)$ and $\rho^*(k)$ (\myfig{sstpower}B,
\myfig{ssttstark}B).

\subsubsection*{Existence of maximal power}

We can explain the existence of a power-maximizing common branch length
$t^*$ under the Jukes-Cantor process, and more generally under any
continuous stationary Markov process, as follows. Fix $k$. At evolutionary
distance zero ($t = 0$), the distribution $p(\mathbf{x}; r)$ in a symmetric
star topology is the same for every mutation rate $r$. Thus the null and
alternative hypotheses coincide. In this circumstance, the power is easily
seen to equal $\alpha$. In the limit of evolutionary time, as $t \to
\infty$, the distribution of each descendant base approaches the process's
stationary distribution, independent of the ancestral base. Since the
stationary distribution does not involve the rate $r$, all conserved and
nonconserved distributions converge to the same limit. The limiting power
in $t$ is therefore again $\alpha$. The fact that power begins at $\alpha$
when $t = 0$ and approaches $\alpha$ as $t \to \infty$, together with the
fact that power is continuous in $t$ and greater than $\alpha$ on
$(0,\infty)$, implies a maximal power must be attained by some finite
$t^*(k)$.

\subsection*{Empirical power analysis}

\begin{figure}
  \centering
  \includegraphics[scale=.89]{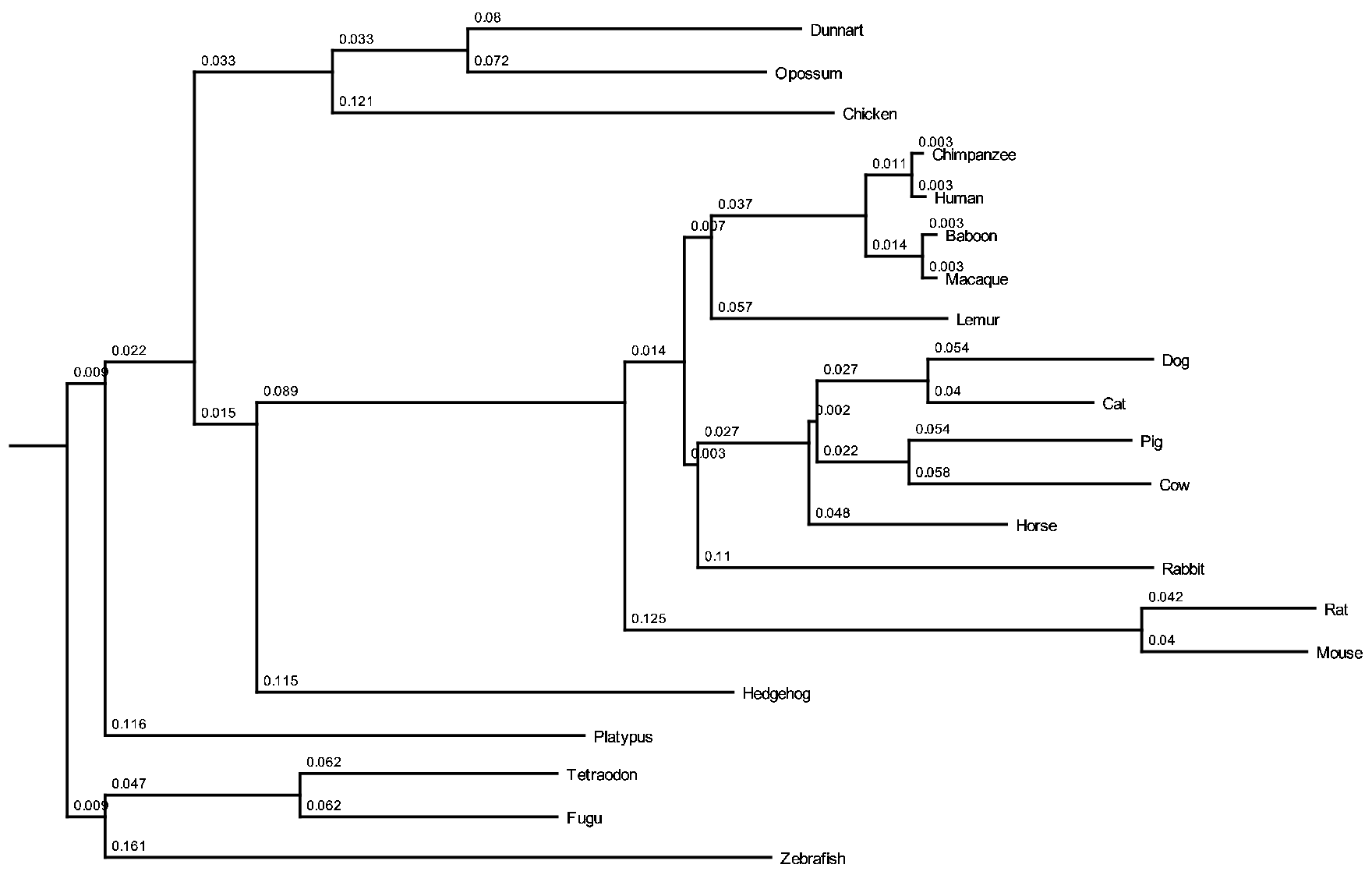}
  \caption{The 21-species phylogenetic tree estimate used in the
    empirical power analysis.}
  \label{fig:tree}
\end{figure}

We constructed a multiple alignment of 21 sequences~(\mytbl{species}) from
the CFTR data set~\cite{ThoEtAl2003} using
\textsc{mavid}~\cite{BraPac2004}. We then used maximum
likelihood~\cite{OlsEtAl1994,Fel1981} to fit a phylogenetic tree topology
and branch lengths~(\myfig{tree}) to to the alignment. Both the phylogeny
estimation and subsequent power analysis employed the nucleotide
substitution process of Felsenstein~\cite{FelChu1996}, using a
transition-transversion ratio of 2:1 and a uniform equilibrium nucleotide
distribution. Branch lengths $\{t_j\}$ are measured in expected number of
substitutions at an exonic aligned site.

The phylogenetic topology of \myfig{tree} differs in a few ways from
estimates based on considerations of large-scale indel mutations and
morphology, for example in its placement of the chicken and platypus. At
issue here, however, is its suitability for a single-site power analysis
under a substitutional mutation model. We chose our tree estimation
procedure to obtain a phylogeny compatible with the data and directed to
this goal.

Finding the $k$-MPSS in a phylogeny is a combinatorial optimization
problem, which we solve directly in small to moderate-sized cases by
evaluating the power of the likelihood-ratio test based on every candidate
Steiner subtree~(\mytbl{stp}). We can also solve the problem directly for
larger $k$, by constraining the species at many of the leaves in the
subtree~(\mytbl{clampedstp}). In order to compute power with a particular
subtree, we used a Monte Carlo strategy. For each setting of $r_N$, with
$\alpha = 0.05$, we generated 100,000 realizations from the null ($r =
r_N$) and alternative ($r = 1$) distributions on the leaves of the full
phylogeny. This induced null and alternative empirical distributions on the
leaves of every possible subtree, from which we obtained approximations to
the true null and alternative distributions of the likelihood-ratio test.
These yielded approximate critical values as well as power estimates. We
repeated the process of simulation and subtree power estimation ten times
for each parameter setting; Tables~\ref{tbl:stp}~and~\ref{tbl:clampedstp}
show averages and standard errors across repetitions.

\bibliography{tech-report}

\clearpage

\begin{table}
  \centering
  \begin{tabular}{rl}
    & Species \\ \hline
    \rule{0ex}{1em} 1 & Baboon \\
    2 & Cat \\
    3 & Chicken \\
    4 & Chimpanzee \\
    5 & Cow \\
    6 & Dog \\
    7 & Dunnart \\
    8 & Fugu \\
    9 & Hedgehog \\
    10 & Horse \\
    11 & Human \\
    12 & Lemur \\
    13 & Macaque \\
    14 & Mouse \\
    15 & Opossum \\
    16 & Pig \\
    17 & Platypus \\
    18 & Rabbit \\
    19 & Rat \\
    20 & Tetraodon \\
    21 & Zebrafish
  \end{tabular}
  \caption{The 21 species whose CFTR region sequence data underlie the
    empirical power analysis.}
  \label{tbl:species}
\end{table}

\clearpage

\begin{table}
  {\centering \small
    \begin{tabular}{r|rlrrr|}

 \multicolumn{1}{c}{}      &      &                                        &              & $t$ vs. & \multicolumn{1}{c}{}     \\
 \multicolumn{1}{c}{$r_N$} & Size & Species: $\star$ = \textsc{mpss}, $\dag$ = \textsc{mdss} & Power\% (SE) & \textsc{mpss}    & \multicolumn{1}{c}{Rank} \\ \cline{2-6}

 \rule{0ex}{1em} & 2 & Rat, Zebrafish $\star \dag$      &  6.79 (0.01) &      &  1.3 \\
    & 3 & Rat, Zebrafish, Chicken $\star \dag$          &  8.30 (0.01) &      &  1.6 \\
  2 & 4 & Rat, Zebrafish, Chicken, Dog $\star \dag$     &  9.61 (0.02) &      &  3.3 \\
    & 5 & Rat, Zebrafish, Chicken, Dog, Dunnart $\star$ & 10.88 (0.03) &      &  4.4 \\
    &   & Rat, Zebrafish, Chicken, Dog, Platypus $\dag$ & 10.80 (0.02) & 2.06 & 21.7 \\ \cline{2-6}

 \rule{0ex}{1em} & 2 & Rat, Zebrafish $\star \dag$        & 10.60 (0.02) &        &    3.2 \\
    & 3 & Rat, Zebrafish, Chicken $\star \dag$            & 21.61 (0.06) &        &    1.8 \\
  5 & 4 & Rat, Zebrafish, Chicken, Dog $\star \dag$       & 39.33 (0.17) &        &    5.2 \\
    & 5 & Rabbit, Cat, Dunnart, Chicken, Hedgehog $\star$ & 49.96 (0.07) &        &   12.2 \\
    &   & Rat, Zebrafish, Chicken, Dog, Platypus $\dag$   & 47.31 (0.07) & 25.82 & 3894.4 \\ \cline{2-6}

 \rule{0ex}{1em} & 2 & Dunnart, Lemur $\star$           & 13.30 (0.03) &        &   21.0 \\
    &   & Rat, Zebrafish $\dag$                         & 12.67 (0.02) &  16.67 &  153.0 \\
    & 3 & Dunnart, Cat, Zebrafish $\star$               & 37.53 (0.11) &        &   10.4 \\
 10 &   & Rat, Zebrafish, Chicken $\dag$                & 36.83 (0.12) &   4.13 &   77.2 \\
    & 4 & Dunnart, Chicken, Hedgehog, Opossum $\star$   & 64.69 (0.05) &        &    4.4 \\
    &   & Rat, Zebrafish, Chicken, Dog $\dag$           & 56.21 (0.06) & 105.70 & 4439.3 \\
    & 5 & Macaque, Lemur, Dog, Cow, Pig $\star$         & 69.75 (0.11) &        &    8.6 \\
    &   & Rat, Zebrafish, Chicken, Dog, Platypus $\dag$ & 66.86 (0.07) &  22.28 & 4867.4 \\ \cline{2-6}
    \end{tabular}}
\caption{The $k$-\textsc{mpss} and $k$-\textsc{mdss} as a
  function of the nonconserved rate $r_N$ and the size $k$ of the subtree,
  with $\alpha = 0.05$ throughout. Results are across 10 repetitions of the
  Monte Carlo power estimation procedure (see Appendix). The last three
  columns display the average power (and standard error), the $t$-statistic
  for the power difference between the $k$-\textsc{mdss} and the
  $k$-\textsc{mpss} (in cases where they differ), and the average power
  ranking (among all subtrees). Since $r_C$ is calibrated to exonic
  conservation, the settings of $r_N$ range from a neutral rate ($r_N =
  2$)~\cite{YapPac2004} towards extreme single-site mutability.}
\label{tbl:stp}
\end{table}

\clearpage

\begin{table}
  {\small \centering
    \begin{tabular}{r|rlrrr|}

 \multicolumn{1}{c}{}      &      &                                            &              & $t$ vs. & \multicolumn{1}{c}{}     \\
 \multicolumn{1}{c}{$r_N$} & Size & New species: $\star$ = \textsc{mpss}, $\dag$ = \textsc{mdss} & Power\% (SE) & \textsc{mpss}    & \multicolumn{1}{c}{Rank} \\ \cline{2-6}

 \rule{0ex}{1em} & 9 & \textit{\{clamped species only\}}         & 12.81 (0.03) &      &     \\
    & 10 & Dunnart $\star$                                       & 14.42 (0.04) &      & 1.1 \\
    &    & Platypus $\dag$                                       & 14.25 (0.04) & 2.92 & 3.4 \\
 2  & 11 & Dunnart, Platypus $\star$                             & 16.08 (0.05) &      & 1.6 \\
    &    & Platypus, Hedgehog $\dag$                             & 15.85 (0.04) & 3.62 & 6.2 \\
    & 12 & Dunnart, Platypus, Hedgehog $\star \dag$              & 17.88 (0.06) &      & 1.5 \\
    & 13 & Dunnart, Platypus, Hedgehog, Rabbit $\star \dag$      & 19.80 (0.08) &      & 1.1 \\
    & 14 & Dunnart, Platypus, Hedgehog, Rabbit, Cow $\star \dag$ & 21.41 (0.08) &      & 1.6 \\ \cline{2-6}

 \rule{0ex}{1em} & 9 & \textit{\{clamped species only\}}         & 56.44 (0.16) &      &      \\
    & 10 & Dunnart $\star$                                       & 65.59 (0.20) &      &  1.0 \\
    &    & Platypus $\dag$                                       & 64.74 (0.17) & 3.18 &  3.0 \\
    & 11 & Dunnart, Opossum $\star$                              & 71.05 (0.09) &      &  2.3 \\
 5  &    & Platypus, Hedgehog $\dag$                             & 70.54 (0.06) & 4.74 & 14.6 \\
    & 12 & Dunnart, Platypus, Hedgehog $\star \dag$              & 72.77 (0.08) &      &  1.2 \\
    & 13 & Dunnart, Platypus, Hedgehog, Rabbit $\star \dag$      & 76.02 (0.13) &      &  1.0 \\
    & 14 & Dunnart, Platypus, Hedgehog, Rabbit, Opossum $\star$  & 80.41 (0.10) &      &  2.2 \\
    &    & Dunnart, Platypus, Hedgehog, Rabbit, Cow $\dag$       & 80.08 (0.14) & 1.88 &  2.1 \\ \cline{2-6}

 \rule{0ex}{1em} & 9 & \textit{\{clamped species only\}}         & 86.61 (0.06) &      &      \\
    & 10 & Platypus $\star \dag$                                 & 91.67 (0.06) &      &  1.3 \\
    & 11 & Dunnart, Opossum $\star$                              & 94.07 (0.02) &      &  3.3 \\
    &    & Platypus, Hedgehog $\dag$                             & 93.96 (0.03) & 2.66 & 10.7 \\
10  & 12 & Dunnart, Platypus, Rabbit $\star$                     & 95.84 (0.03) &      &  2.4 \\
    &    & Dunnart, Platypus, Hedgehog $\dag$                    & 95.79 (0.30) & 1.30 &  4.4 \\
    & 13 & Dunnart, Platypus, Rabbit, Opossum $\star$            & 97.31 (0.02) &      &  4.6 \\
    &    & Dunnart, Platypus, Rabbit, Hedgehog $\dag$            & 97.29 (0.02) & 0.85 &  6.6 \\
    & 14 & Dunnart, Platypus, Rabbit, Hedgehog, Opossum $\star$  & 97.99 (0.01) &      &  2.4 \\
    &    & Dunnart, Platypus, Rabbit, Hedgehog, Cow $\dag$       & 97.95 (0.02) & 1.83 &  7.6 \\ \cline{2-6}

    \end{tabular}}
  \caption{The $k$-\textsc{mpss} and $k$-\textsc{mdss}, under the
    constraint that the following nine species are included in the subtree:
    human, mouse, rat, chimpanzee, dog, chicken, fugu, zebrafish, and
    tetraodon. The scheme of the table is the same as Table 1.}
  \label{tbl:clampedstp}
\end{table}

\clearpage

\clearpage

\clearpage

\end{document}